\documentclass[aps,pra,superscriptaddress,amsmath,amssymb,preprintnumbers,showpacs,twocolumn]{revtex4}
\usepackage{graphicx,color}
\usepackage{amssymb}
\usepackage{amsmath}
\usepackage{bm}
\newcommand{\tr}{\mathop{\mathrm{tr}}\nolimits}    \newcommand{\HA}{\mathop{\mathcal{H}}\nolimits} 
 \newcommand{\LA}{\mathop{\mathcal{L}}\nolimits}   \newcommand{\SA}{\mathop{\mathcal{S}}\nolimits}
   \newcommand{\R}{\mathop{\mathbb{R}}\nolimits}  \newcommand{\N}{\mathop{\mathbb{N}}\nolimits}
\newcommand{\I}{\mathop{\mathbb{I}}\nolimits}   

\newtheorem{Thm}{Theorem} \newtheorem{Prop}{Proposition} \newtheorem{Lem}{Lemma}    \newtheorem{Rem}{Remark}    

\newcommand{\bra}[1]{\langle #1 |}
\newcommand{\ket}[1]{| #1 \rangle}
\newcommand{\bracket}[2]{\langle #1 | #2 \rangle}
\newcommand{\ketbra}[2]{| #1 \rangle \langle #2 | }

\definecolor{dgreen}{rgb}{0,0.5,0}

\definecolor{delete}{cmyk}{0.5,0,0,0}
\definecolor{deletey}{cmyk}{0,0.5,0,0}

\begin{document}


\title{Possibility of a minimal purity-measurement scheme critically depends on the parity of dimension of the quantum system}


\author{T. Tanaka}
\affiliation{Department of Physics, Waseda University, Tokyo 169-8555, Japan}

\author{G. Kimura}
\affiliation{College of Systems Engineering and Science, Shibaura Institute of Technology, Tokyo 108-8548, Japan}

\author{H. Nakazato}
\affiliation{Department of Physics, Waseda University, Tokyo 169-8555, Japan}


\begin{abstract}

In this paper, we investigate the possibility of measuring the purity of a quantum state (and the overlap between two quantum states) within a minimal model where the measurement device is minimally composed. 
The minimality is based on the assumptions that (i) we use a yes-no measurement on a single system to determine the single value of the purity in order not to extract other redundant information, and (ii) we use neither ancilla nor random measurement.
We show that the measurability of the purity within this model critically depends on the parity of dimension of the quantum system: 
the purity measurement is possible for odd dimensional quantum systems, while it is impossible for even dimensional cases.  
\end{abstract}


\maketitle

\section{introduction}\label{intro}
There are several functionals of quantum states each of which represents physical or informational meanings. 
One of the most important examples would be the von Neumann entropy \cite{ref:vN}, which has some operational meanings, e.g., 
of the optimal compression rate \cite{ref:OptimalRate} or of the optimal distilability of entanglement for the reduced state from a pure composite state \cite{ref:EntDis}. 
To investigate the experimental determination or the estimation of such functionals for unknown quantum states is one of the most important issues, especially in the field of quantum information science \cite{NC,ref1,ref2}. 

One can calculate the value of any functional for a quantum state after the construction of the state through the quantum state tomography. 
However, this approach requires extra information to determine the single value of the functional besides the computational resources to calculate it, and thus is extremely inefficient.
Indeed, the state of the $D$-dimensional quantum system includes $D^2 -1$ independent parameters and  the number of them quadratically increases as the dimension of the quantum system becomes large. 
(Throughout this paper, $D$ represents the dimension of the Hilbert space $\HA$ associated to a quantum system.) 
Therefore, the direct determination of functionals for unknown quantum states is desirable from the view point of the efficiency. 
Moreover, this would also provide an operational meaning of the functional.  

In this paper, we focus on the purity, perhaps the simplest one among useful functionals for quantum states, 
and investigate its direct measurability. 
The purity of a quantum state described by a density operator $\rho$ is defined by 
\begin{equation}\label{eq:Purity}
P(\rho):= \tr \rho^2 .
\end{equation}
It is usually considered as a natural measure of ``pureness" (``coherence") of quantum states. 
Indeed, the quantum state is pure iff its purity takes the maximum value $1$, while it is the maximally mixed state $\rho_{m}:= \I/D$ iff its purity takes the minimum value $1/D$. 
In two-qubit systems, it is well known that the purity of the reduced state has a direct relation to the concurrence \cite{ref:WoottersConc,ref:Conc}. 
The closely related quantities include the linear entropy $E(\rho):= 1 - P(\rho)$, and the overlap \cite{ref:flip,ref:ekert} between two states described by $\rho$ and $\sigma$:  
\begin{equation}\label{eq:Overlap}
O(\rho,\sigma):= \tr \rho \sigma.  
\end{equation}
If one of $\rho$ and $\sigma$ is a pure state, it can be written as $O(\rho,\sigma) = F(\rho,\sigma)^2$ with the fidelity $F(\rho,\sigma) := \tr \sqrt{\sqrt{\rho}\sigma \sqrt{\rho}}$ \cite{tomography}. 
As the overlap is a simple generalization of the purity: $P(\rho) = O(\rho,\rho)$, we also discuss its direct measurability along with that of the purity. 
Note that the device to measure the overlap enables us to measure the purity, while the impossibility of the purity measurement implies that of the overlap measurement (See Sec.~\ref{sec:main}). 

There are several methods to measure the purity $P(\rho)$ and the overlap $O(\rho, \sigma)$ without resort to the state tomography. 
In Refs.~\cite{ref:flip,ref:ekert}, a circuit for measuring the overlap is presented on the basis of an interferometric setup and controlled unitary operations. 
The value of the overlap is transfered to an interference pattern in an ancilla system.   
An alternative scheme for measuring the purity is presented without resorting either to interferometry or to controlled unitary operations \cite{ref:NTYFP}. 
On the other hand, the random-measurement method is proposed for the purity measurement which does not use any ancilla system \cite{random}. 
To perform this scheme, however, one needs to realize random unitary operations distributed according to the Haar measure. 

The purpose of this paper is to investigate the direct determination of the purity of a quantum state (and the overlap between two quantum states) in an optimal way such that the measurement device is minimally composed. 
In particular, we only use a unitary interaction and a local yes-no measurement on a single system without resorting either to ancilla systems or to random measurements (See Sec.~\ref{sec:MinimalModel}).  
Under these restrictions, we theoretically discuss the possibility of the determination of the purity (and the overlap) in a general quantum system. 
Interestingly, the result depends on the dimension of the associated Hilbert space, i.e., the measurement is possible/impossible in odd/even dimensional cases. 

This paper is organized as follows. 
In Sec.~\ref{sec:MinimalModel}, we explain the meaning of the minimality of the purity measurement (and the overlap). 
In Sec.~\ref{sec:swap}, some general properties of the swap operator and (anti)symmetric subspaces of Hermitian operators are presented. 
In Sec.~\ref{sec:main}, we show our main results about the possibility of measuring the purity of any unknown quantum state (and the overlap between two quantum states) within the minimal model. 
In Sec.~\ref{sec:ex}, we give examples of a measurement operator and a unitary gate to measure the purity (and the overlap) in the minimal model for odd dimensional quantum systems.
Section~\ref{s_and_d} is devoted to the summary and discussion.

\section{Minimal model of purity-measurement}\label{sec:MinimalModel}

\begin{figure}[tb]
\includegraphics[width=85mm]{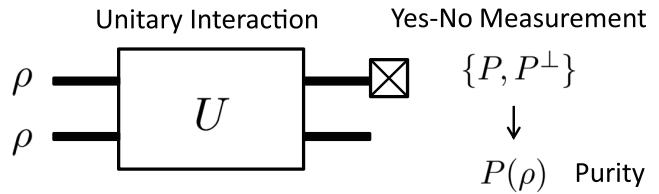}
\caption{Minimal model for the purity measurement is composed of a yes-no PVM measurement $\{P,P^\perp \}$ on a single system after the unitary interaction $U$ on the input $\rho \otimes \rho$. }\label{fig:MinModel}
\end{figure}

Let us start from clarifying the  meaning of the minimality of the model for the purity measurement. 
We assume as usual that an i.i.d. (identical and independent distribution) of unknown quantum states is available. 
To compose the model as minimally as possible, 
we use neither ancilla systems nor random measurements, and thus all elements available for this model are a unitary interaction and a PVM (projection valued measure) measurement \footnote{Recall that the $n$-valued PVM (projection valued measure) is an $n$-tuples of projection operators $(P_i)_{i=1}^n$ with the completeness condition $\sum_i P_i = \I$, which corresponds to the measurement of an observable with the eigen-projection $P_i$. }. That is, we only utilize standard ingredients of quantum mechanics \cite{ref:Dirac}. 
In the following analysis, we also consider a POVM (positive operator valued measure) measurement as a possible generalization, but it turns out that the result does not change at all. 
In particular, the measurement should be a yes-no measurement (i.e., two-valued measurement) for the purpose of the direct determination of the purity from the single output imformation. 
In this setting, it is impossible to measure the purity using a single system as is easily shown from the linearity of the probability on the density operator.  
Therefore, we use a bipartite system $\HA \otimes \HA$ with the input of two copies of a quantum state $\rho \otimes \rho$ \footnote{Note that if we are allowed to perform a POVM measurement with more than two outputs, it is possible to perform the state tomography and thus the purity also can be calculated from the measurements (See also \cite{ref:Brun}). 
Note also that if we are allowed to perform the random measurement, then it is possible to determine the purity by averaging the square of probabilities in each random measurement \cite{random}. 
However, as is stated in the introduction, these are inefficient ways and not the case we want.} 
and the simplest model for the purity measurement would be composed of a unitary interaction $U$ on $\rho \otimes \rho$ 
followed by a local yes-no measurement $\{P,P^\perp := \I - P \}$ (with a projection operator $P$ on $\HA$) on a single system. 
By performing the final PVM measurement, one obtains the probability to have ``yes" (or alternatively ``no") outputs: 
\begin{equation}\label{eq:prob}
{\rm Pr}(\rho) := \tr[ (P \otimes \I) U \rho\otimes \rho U^\dagger],
\end{equation}
from which one would determine the value of the purity $P(\rho)$. 
We call these series of procedures the minimal model for the purity measurement (See Fig.~\ref{fig:MinModel}). 
\begin{Rem}\label{rem:OV}
We also define the minimal model for the overlap measurement just by replacing the input $\rho \otimes \rho$ in the minimal model of the purity measurement with the input $\rho \otimes \sigma$.  
\end{Rem}

At this stage, one might wonder whether the probability \eqref{eq:prob} with a suitable choice of an interaction and a measurement becomes the purity itself. 
However, a short consideration reveals that it is impossible. 
Let us show this for the easiest case of a PVM measurement of rank $1$. (The general cases will be treated rigorously in the next section.)  
Let $U$ be a unitary operator on $\HA \otimes \HA$ for the interaction and let $P = \ketbra{\phi}{\phi}$ be a projection operator of rank $1$ on $\HA$ for a yes-no measurement. 
Let $\{\ket{i}\}_{i=1}^D$ be an ONB (orthonormal basis) of $\HA$ such that $\ket{1} = \ket{\phi}$, then $\ket{\psi_{ik}} := U \ket{i} \ket{k}$ is an ONB of $\HA \otimes \HA$. 
 With these bases, the purity and the probability \eqref{eq:prob} are written as  
\begin{equation}\label{eq:PurityProb}
P(\rho) = \sum_{i,j}p_{ij}p_{ji}, \ {\rm Pr}(\rho) = \sum_{i,j,k,l=1}^D p_{ij} p_{kl} \bracket{\psi_{ik}^{(1)}}{\psi_{jl}^{(1)}}
\end{equation}
where $p_{ij} : =\bra{i}\rho\ket{j}$ and $\ket{\psi_{ik}^{(m)}} := \bracket{m}{\psi_{ik}}$. 
To make the purity equal to the probability for any state $\rho$, one has to choose $U$ such that 
$\bracket{\psi_{ik}^{(1)}}{\psi_{jl}^{(1)}} = \delta_{il}\delta_{jk}$. 
However, it is easy to see that there are no such ONB $\{\ket{\psi}_{ik}\}_{i,k=1}^D$ of $\HA\otimes \HA$ 
\footnote{
Let $\{\ket{\psi_{ik}}\}_{i,k=1}^D$ of $\HA \otimes \HA$ satisfy the condition $\bracket{\psi_{ik}^{(1)}}{\psi_{jl}^{(1)}} = \delta_{il}\delta_{jk}$ where $\sum_{m=1}^D \ket{m}\ket{\psi_{ik}^{(m)}} := \ket{\psi_{ik}}$. 
For $i =l \not = k=j$, this condition and the Schwartz inequality yield $1 \leq \| \psi _{ij}^{(1)}\| ^2 \| \psi _{ji}^{(1)} \| ^2 \leq (\sum_{m=1}^{D} \| \psi_{ij}^{(m)} \| ^2)(\sum_{n=1}^{D} \| \psi_{ji}^{(n)} \| ^2) =1$. Therefore, $\| \psi _{ij}^{(m)}\| = \| \psi _{ji}^{(m)}\| =0 (m=2,\dots,D)$ and thus we have $ \ket{\psi_{ij}} = \ket{1} \ket{\psi_{ij}^{(1)}}$ and $\ket{\psi_{ji}} = \ket{1} \ket{\psi _{ji}^{(1)}}$. 
Hence, $\{\ket{\psi}_{ik}\}_{i,k=1}^D$ cannot be an ONB.}. 

Therefore, to determine the purity of any unknown state from the probability \eqref{eq:prob}, one needs a suitable function $f$ which relates the probability to the purity by 
\begin{equation}\label{eq:cond}
P(\rho) = f({\rm Pr}(\rho)), \ \forall \rho \in \SA(\HA).
\end{equation}
Here, $\SA(\HA)$ denotes the set of all density operators on $\HA$. 
One might anticipate the extra computational resource for $f$, but it turns out to be sufficient to consider a linear function of the form $f(t) = a t + b \ (a \neq 0,b \in \R)$: 
 \begin{Prop}\label{prop:linearity}
(Linearity of function $f$) The existence of a function $f$ satisfying Eq.~\eqref{eq:cond} is equivalent to 
\begin{equation}\label{eq:linearity}
\exists x \not= 0,y \in \R \ {\rm s.t.} \ {\rm Pr}(\rho) =  x P(\rho) + y, \ \forall \rho \in \SA(\HA).
\end{equation}
\end{Prop}
(See Appendix \ref{sec:linearity} for the proof.)

To summarize, the minimal model for the purity measurement is composed of a unitary interaction on $\rho \otimes \rho$ followed by a yes-no PVM measurement 
such that the linear transformation of the ``yes"-probability yields the purity (See Fig.~\ref{fig:MinModel}).

\section{Swap operator and subspace}\label{sec:swap}

The swap operator is a key tool for the analysis of the purity measurement. 
Indeed, it is widely known that the purity is written as  
\begin{equation}\label{eq:PS}
P(\rho) = \tr (S \rho \otimes \rho),
\end{equation}
where $S$ is the swap operator on $\HA\otimes \HA$ defined by 
\begin{equation}
S \ket{\psi} \otimes \ket{\phi} := \ket{\phi} \otimes \ket{\psi} , \ ^\forall \ket{\psi}, \ket{\phi} \in \HA .
\end{equation}
That is, the measurement of the swap operator $S$ (an observable) makes it possible to obtain the purity as an expectation value of $S$ under the state $\rho \otimes \rho$. 
However, mostly because $S$ is a global operator, this strategy is not applicable to the minimal model for the purity measurement. 
Notice that the overlap is also given by 
\begin{equation}\label{eq:OvS}
O(\rho,\sigma) = \tr (S \rho \otimes \sigma). 
\end{equation}

In the following, we show some general properties on the swap operator and (anti)symmetric subspaces of Hermitian operators for later convenience. 
In fact, the number of the dimension of (anti)symmetric subspaces plays an important role when we discuss the measurability of the purity within the minimal model.
Let $\LA ^h(\HA)$ be the set of all the Hermitian operators on $\HA$. 
Note that $\LA ^h(\HA )$ is a $({\rm dim}(\HA))^2$ dimensional real Hilbert space with respect to the Hilbert-Schmidt inner product. 
(In the following, we always use the Hilbert-Schmidt inner product for operators.)  

\begin{Lem}\label{lem:Swap}
$S^2 = \I \otimes \I$ and $\tr S = D$; thus eigenvalues of $S$ are $\pm 1$ with the multiplicities $\frac{D(D\pm 1)}{2}$, respectively. 
\end{Lem}
{\bf Proof}.
It is trivial to show $S^2 = \I \otimes \I$ and since $S$ is both unitary and Hermitian, the eigenvalues of $S$ are $\pm 1$. 
From the definition of $S$, $S \I \otimes \I = \sum_{i,j=1}^D \ketbra{ji}{ij}$ for an ONB $\{\ket{i}\}_{i=1}^D$, and 
by taking the trace of this equation, we have $\tr S = D$. 
\hfill $\blacksquare$

Using the swap operator, one can define symmetric and antisymmetric subspaces of Hermitian operators as follows:
\begin{equation}\label{eq:SAS}
\begin{split}
V_{sym}^h \ &:= \ \{\ A \in \LA ^h (\HA \otimes \HA ) | \Lambda (A) = A\}, \\
V_{asym}^h \ &:= \ \{\ A \in \LA ^h(\HA \otimes \HA ) | \Lambda (A) = -A\},
\end{split}
\end{equation}
where $\Lambda$ is the swap map on $\LA^h(\HA \otimes \HA)$ defined by 
\begin{equation}
\Lambda (A):= S A S, \ A \in \LA ^h (\HA \otimes \HA).   
\end{equation}
As the following Lemma shows, these subspaces form an orthogonal decomposition of $\LA^h(\HA\otimes \HA)$ \footnote{
This result itself is an immediate consequence from the fact that $\Lambda$ is a Hermitian (super)operator on $\LA^h(\HA\otimes \HA)$ whose 
eigenvalues are $\pm 1$ with eigenspaces $V_{sym}^h$ and $V_{asym}^h$.}: 
\begin{Lem}\label{lem:orthoSymAssym}
$\LA^h(\HA\otimes \HA) = V_{sym}^h \oplus V_{asym}^h$ where ${\rm dim}(V_{sym}^h)=\frac{D^2(D^2+1)}{2}$ and ${\rm dim}(V_{asym}^h)=\frac{D^2(D^2-1)}{2}$.
\end{Lem}
{\bf Proof}. For any $A \in V_{sym}^h$ and $B \in V_{asym}^h$, as $S^2 = \I \otimes \I$, we have 
$\tr(A^{\dagger}B) = \tr(SA^{\dagger}SSBS)= \tr(\Lambda(A^{\dagger}) \Lambda(B))=-\tr(A^{\dagger} B)$ where we have used the definition \eqref{eq:SAS}. 
This shows that $V_{sym}^h$ and $V_{asym}^h$ are orthogonal with respect to the Hilbert-Schmidt inner product. 
In order to show $\LA^h(\HA\otimes \HA) = V_{sym}^h \oplus V_{asym}^h$ with $\rm{dim}(V_{sym}^h)=\frac{D^2(D^2+1)}{2}$ and $\rm{dim}(V_{asym}^h)=\frac{D^2(D^2-1)}{2}$, 
we construct the corresponding bases from an arbitrary ONB $\{X_i\}_{i=1}^{D^2}$ of $\LA^h(\HA)$ as follows:  
one can take the basis of $V_{sym}^h$ from $\frac{D^2(D^2+1)}{2}$ numbers of operators:   
\begin{equation}\label{eq:bsym}
 \{X_i \otimes X_i \}\ _{i=1}^{D^2} \ \rm{and} \ \{\frac{1}{\sqrt 2}(X_i \otimes X_j + X_j \otimes X_i) \}\ _{1 \leq i < j \leq D^2},
\end{equation}
and the basis of $V_{asym}^h$ from $\frac{D^2(D^2-1)}{2}$ numbers of operators:
\begin{equation}
\{\frac{1}{\sqrt 2}(X_i \otimes X_j - X_j \otimes X_i) \}\ _{1 \leq i < j \leq D^2}. 
\end{equation}
It is easy to show that these $D^4$ numbers of operators form a linearly independent set and thus are complete in $\LA^h(\HA \otimes \HA)$.  
\hfill $\blacksquare$
\begin{Lem}\label{lem:ACom} For any $A \in V^h_{asym}$ and $k \in \N$, 
\begin{eqnarray*}
&\rm{(i)}& \tr A^{2k-1} = 0, \\
&\rm{(ii)}& \tr A^{k} S = 0, \\
&\rm{(iii)}& A \ \rm{anticommutes \ with} \ S. 
\end{eqnarray*}
\end{Lem}
{\bf Proof}. 
(i) We have $\tr A^{2k-1} = \tr (S A S)^{2k-1} = \tr (-A)^{2k-1} = -\tr A^{2k-1}$, where 
in the first equality we have used $S^2 = \I \otimes \I$ and the cyclic property of the trace and in the second equality $S A S = \Lambda(A) = -A $. 
Thus $\tr A^{2k-1} = 0$. 
(ii) Similarly, $\tr A^{k}S = \tr A^{k-1}S S A S = \tr A^{k-1}S (-A) = - \tr A^{k}S$.
(iii) Multiplying $S$ to $-A = SA S$, we have $ - A S = SA  S S = S A$.

\hfill $\blacksquare$

The following proposition is critical for the measurability of the purity within the minimal model and would be useful when one
considers other measurement problems:  
\begin{Prop}\label{prop:prod}For any $A \in \LA ^h(\HA \otimes \HA)$,
\begin{eqnarray*}
&\rm{(i)}& \  \tr(A \rho \otimes \sigma) = 0, \ ^\forall \rho , \sigma \in \SA(\HA) \ \Leftrightarrow A = 0, \\
&\rm{(ii)}& \ \tr(A \rho \otimes \rho) = 0, \ ^\forall \rho \in \SA(\HA) \ \Leftrightarrow A \in V_{asym}^h.
\end{eqnarray*}
\end{Prop}
To prove this, we first show the following Lemma. In what follows, we mean by span the real linear span: 
\begin{equation}
{\rm span}\  X:= \{\sum_i a_i x_i \ | \ a_i \in \R, x_i \in X \}. 
\end{equation}
\begin{Lem}\label{lem:sps}(Spans of Product States)
\begin{eqnarray*}
& \rm{(i)}& {\rm span} \{\ \rho \otimes \sigma \ | \ \rho,\sigma \in \SA(\HA) \}\ = \LA^h(\HA \otimes \HA) \\ 
& \rm{(ii)} & {\rm span} \{\ \rho \otimes \rho \ | \ \rho \in \SA(\HA) \}\ = V_{sym}^h
\end{eqnarray*}
\end{Lem}
{\bf Proof}. 
(i) It is enough to show $A \otimes B \in \rm{span} \{\ \rho \otimes \sigma \ | \ \rho,\sigma \in \SA(\HA) \}\ $ for any $A,B \in \LA^h(\HA)$. 
Let $A_\pm$ ($B_\pm$) be positive and negative parts of $A$ ($B$) so that $A = A_+ - A_-$ ($B = B_+ - B_-$) \footnote{Note that a Hermitian operator $A$ is divided into a part with positive eigenvalues $A_{+}$ and a part with negative eigenvalues $-A_{-}$, i.e., $A = A_{+} - A_{-}$, where
$A_{\pm}$ are positive operators. The part $A_{+}(A_{-})$ is referred to as the positive (negative) part.}. 
Observe the following identity: 
\begin{equation}
\begin{split}
A \otimes B = A_{+} \otimes B_{+} - A_{-} \otimes B_{+} - A_{+} \otimes B_{-} + A_{-} \otimes B_{-}. 
\end{split}
\end{equation}
As the right hand side is a real linear span of the products of density operators, this completes the proof of (i). 

(ii) The inclusion $\subseteq $ holds as $\Lambda (\rho \otimes \rho ) = \rho \otimes \rho$. 
To prove the opposite relation, we show that a basis of $V_{sym}^h$ is in $\rm{span}\{\ \rho \otimes \rho \ | \ \rho \in \SA(\HA) \}$. 
From Eq.~\eqref{eq:bsym}, it is enough to show this for the operator $X$ of the form  
\begin{equation}
 X = A \otimes B + B \otimes A, \ A,B \in \LA^h(\HA) . 
\end{equation}
Moreover, since $X$ can be written as 
\begin{equation}
X = \frac{1}{2}((A+B)\otimes (A+B) - (A-B)\otimes (A-B)),
\end{equation}
it is sufficient to show that an operator of the form $C \otimes C$ is a real linear span of products of identical density operators. 
This follows from the identity: 
\begin{equation}
C \otimes C = 2 C_+ \otimes C_+ + 2 C_- \otimes C_- - (C_+ + C_-)\otimes (C_+ + C_-),
\end{equation}
where $C_\pm$ are the positive and negative parts of $C$. 
This completes the proof of (ii). \hfill $\blacksquare$

{\bf Proof of Proposition \ref{prop:prod}}. 
(i) Let $\tr(A \rho \otimes \sigma) = 0$ for any $\rho , \sigma \in \SA(\HA)$. 
Then from Lemma \ref{lem:sps}-(i), $A \in \LA ^h(\HA \otimes \HA) ^{\perp } =\{0\}$. 
(ii) Let $\tr(A \rho \otimes \rho) = 0$ for any $ \rho \in \SA(\HA)$. 
Then, from Lemma \ref{lem:orthoSymAssym} and Lemma \ref{lem:sps}-(ii), $A \in (V_{sym}^h)^{\perp } = V_{asym}^h$. 
\hfill $\blacksquare$

\section{Measurability of purity}\label{sec:main}

In this section, we discuss the measurability of the purity within the minimal model given in the previous sections. 
Interestingly, the result is critically dependent on the parity of dimension of the associated Hilbert space: 
\begin{Thm}\label{thm:main}
The measurement of the purity (of any unknown state) within the minimal model is possible for odd dimensional Hilbert spaces, 
while it is impossible for even dimensional cases. 
\end{Thm}
Before giving a proof of Theorem \ref{thm:main}, several remarks are in order, the proofs of which are included in that of Theorem \ref{thm:main}.
\begin{Rem}\label{rem:POVMcase}
The results do not change even using a two-valued POVM measurement as the final measurement. 
\end{Rem}
\begin{Rem}\label{rem:LocalTwoBody}
If one utilizes a global measurement, the purity measurement is possible for any dimensional case (e.g., using the measurement of the swap operator) with a unitary interaction. 
Moreover, one can measure the purity with a local two-body measurement \rm{\footnote{The local two-body measurement is a measurement on the bipartite system in $\HA \otimes \HA$ described by a measurement operator$P \otimes Q$, where $P$ and $Q$ are projection operators on $\HA$.}}. 
\end{Rem}
In addition, the same result holds for the overlap measurement (See Remark \ref{rem:OV}): 
\begin{Prop}\label{prop:Overlap}
The measurement of the overlap (between any unknown states) within the minimal model is possible for odd dimensional Hilbert spaces, 
while it is impossible for even dimensional cases (e.g., n-qubit systems). 
\end{Prop}

\begin{Rem}\label{rem:InfDim}
In the case of infinite dimension, the measurements of the overlap and thus the purity are possible within the minimal model. 
\end{Rem}

{\bf Proof of Theorem \ref{thm:main}}. 
Notice that if there exists a minimally composed device which allows us to determine the value of the overlap $O(\rho , \sigma)$ between any states $\rho$ and $\sigma$, the device enables us to measure the purity with the state $\sigma$ replaced by the state $\rho$. On the other hand, if there does not exist a minimal device for measuring the purity $P(\rho) = O(\rho, \rho)$, the ovelap between two states, of course, cannot be determined within the minimal model. 
Bearing this in mind, the proof is organized as follows.
(a) We first prove the former part of Proposition \ref{prop:Overlap} for odd dimensional cases, 
which implies the former part of Theorem \ref{thm:main}. This also covers the proof of Remark \ref{rem:LocalTwoBody}. 
Moreover, we give a sketch of the proof of the Remark \ref{rem:InfDim} for the infinite dimensional case at the end of the part (a).
(b) Next, we prove the latter part of Theorem \ref{thm:main} for even dimensional cases, 
which implies the latter part of Proposition \ref{prop:Overlap}.

(a) We first find a unitary operator $U$ on $\HA \otimes \HA$, a local yes-no PVM measurement $\{P\otimes Q,\I-P\otimes Q\}$ on two-body systems
and a final linear transformation for the overlap measurement (for the minimal model, $Q = \I$).  
Substituting $\mathrm{Pr}(\rho):= \tr [(P\otimes Q) U \rho\otimes \sigma U^\dagger]$ and Eq.~\eqref{eq:OvS} into Eq.~(\ref{eq:linearity}), 
the measurability of the overlap is equivalent to the existance of $x \neq 0, y \in \R$ satisfying
\begin{equation}
\tr [(P\otimes Q) U \rho\otimes \sigma U^\dagger] = x \tr (S \rho \otimes \sigma) + y, \ \forall \rho,\sigma \in \SA(\HA).
\end{equation} 
From Proposition \ref{prop:prod}-(i), moreover, this is equivalent to   
\begin{equation}
U^{\dagger} P \otimes Q U = x S + y \I \otimes \I, \label{eq4}
\end{equation}
where we have used the facts (i) $1 = \tr ((\I \otimes \I) \rho \otimes \sigma)$ and (ii) the cyclic property of the trace.  
Since $P\otimes Q$ and $x S + y\I \otimes \I$ are Hermitian (and thus diagonalizable), it is enough to find $P,Q,x$ and $y$ such that 
$P\otimes Q$ and $x S + y\I \otimes \I$ have the same eigenvalues and multiplicities
\footnote{If Hermitian operators $A$ and $B$ have the same eigenvalues and the multiplicities, then 
eigenvalue decompositions of $A$ and $B$ have the same form: $A = \sum_i \lambda_i \ketbra{a_i}{a_i}$ and 
$B =\sum_i \lambda_i \ketbra{b_i}{b_i}$ where $\lambda_i$ are the (common) eigenvalues and $\{\ket{a_i}\}$ ($\{\ket{b_i}\}$) is an ONB of eigenvectors of $A$ ($B$). 
By choosing a unitary operator as $U = \sum_i \ketbra{a_i}{b_i}$, we have $B = U^\dagger A  U $. 
}. 

Let $k$ and $l$ be the ranks of $P$ and $Q$. Then the left hand side of Eq.~\eqref{eq4} has eigenvalues $1$ and $0$ with the multipliticies $kl$ and $D^2 -kl$, respectively. 
From Lemma \ref{lem:Swap}, on the other hand, the right hand side of Eq.~\eqref{eq4} has eigenvalues $x+y$ and $-x+y$, whose multiplicities are $\frac{D(D+1)}{2}$ and $\frac{D(D-1)}{2}$, respectively. 
Therefore, for the consistency in their eigenvalues and multiplicities on both sides of Eq.~\eqref{eq4}, 
one can choose $k$, $l$, $x$ and $y$ as $\frac{D}{2}$, $D-1$, $\frac{-1}{2}$ and $\frac{1}{2}$, respectively for even dimensional cases. 
On the other hand, for odd dimensional cases, one can choose $k$, $l$, $x$ and $y$ as $\frac{D+1}{2}$, $D$, $\frac{1}{2}$ and $\frac{1}{2}$, respectively, and thus $Q = \I$. 
Therefore, we have shown that the overlap (and thus the purity) are measurable with a local two-body measurement $P \otimes Q$ (Remark \ref{rem:LocalTwoBody}), and especially in the case of odd dimensional cases within a minimal model (the former parts of Theorem \ref{thm:main} and Proposition \ref{prop:Overlap}). 

A similar argument about the measurability of the overlap for the infinite dimensional case follows, 
just by noting the ideal property of a trace class operator in the space of bounded operators \cite{ref:Sch}: for any trace class operator $A$ and for any bounded operator $B$, $A B$ and $B A$ are trace class operators. 
Since $\rho \otimes \sigma$ is a trace class operator and $S,U,P,\I$ are all bounded operators, 
it is enough to find a projection operator $P$ and a unitary operator $U$ satisfying Eq.~\eqref{eq4} with $Q = \I$ and $x = y = \frac{1}{2}$ in order to show the measurability of the overlap within the minimal scheme. 
As $S$ has point spectra $\pm 1$ with infinite dimensional eigenspaces, one can use any projection operator $P$ such that both $P$ and $\I - P$ have infinite dimensional ranges. Then, 
there exists a unitary operator $U$ which connects the eigenspaces of $\frac{1}{2}(S + \I\otimes\I)$ and $P \otimes \I$ (Remark \ref{rem:InfDim}).

(b) Next, we show the latter parts of both Theorem \ref{thm:main} and Proposition \ref{prop:Overlap} about the impossibility of measuring the purity or the overlap within the minimal model for even dimensional cases. 
(Recall that the impossibility of measuring the purity implies that of measuring the overlap. Therefore, in the following, we prove the latter part of Theorem \ref{thm:main}.) 
We show this by contradiction: we assume that with a unitary operator $U$ on $\HA \otimes \HA$, a projection operator $P$ on $\HA$, and $x \neq 0, y \in \R$, Eq.~\eqref{eq:linearity} is satisfied. 
In particular, in order to include Remark \ref{rem:POVMcase}, we replace $P$ with a POVM element $0 \le E \le \I$ in what follows. 
Using $\mathrm{Pr}(\rho):= \tr [E\otimes \I U \rho\otimes \rho U^\dagger]$ and Proposition \ref{prop:prod}-(ii) in Eq.~(\ref{eq:linearity}), this assumption is equivalent to 
\begin{equation}\label{eq5}
U^{\dagger} (E -y\I)\otimes \I U = xS + A_{asym},
\end{equation} 
for some $A_{asym} \in V^h_{asym}$. 
Let $\vec \lambda = (\lambda_1 , \dots, \lambda_D)$ be eigenvalues of $E - y \I$. 
From Eq.~\eqref{eq5}, they are also eigenvalues of $xS + A_{asym}$. 
From now, we show the following two facts about the eigenvalues: 
\begin{equation}\label{eq:lambda>x}
|\lambda_i| \geq |x|, \ \forall i \in \{\ 1,\ldots,D \}\ ,
\end{equation}
and 
\begin{equation}\label{eq6}
\sum_{i=1}^D \lambda_i ^{2k-1} = x^{2k-1}, \forall k \in \N . 
\end{equation}
To show Eq.~\eqref{eq:lambda>x}, let $\ket{\phi_i} \in \HA$ be an eigenvector of $xS + A_{asym}$ with the corresponding eigenvalue $\lambda_i$. 
Note that from Lemma \ref{lem:ACom}-(iii) and $S^2 = \I \otimes \I$,  
\begin{equation}\label{eq:SAsym}
(xS + A_{asym})^2 = x^2 \I \otimes \I + A^2_{asym}, 
\end{equation}
from which we obtain $A^2_{asym} \ket{\phi_i} = (\lambda_i^2 - x^2) \ket{\phi_i}$. 
This means that $\lambda_i^2 - x^2$ is one of eigenvalues of $A^2_{asym}$, which is a positive operator and therefore $\lambda_i^2 \ge x^2$. 
To show Eq.~\eqref{eq6}, we take the trace of Eq.~\eqref{eq5} to the $(2k-1)$-th power and obtain $D \sum_{i=1}^D \lambda_i ^{2k-1}  = \tr (xS + A_{asym})^{2k-1}$. 
However, from Eq.~\eqref{eq:SAsym} we have $\tr (xS + A_{asym})^{2k-1} = \tr (x^{2}\I \otimes \I + A^2_{asym})^{k-1} (xS + A_{asym}) = \tr x^{2k-1} S = D x^{2k-1}$, 
where we have used Lemma \ref{lem:ACom}-(i),(ii) in the third equality. 

Finally, we show that the conditions \eqref{eq:lambda>x} and \eqref{eq6} are not compatible if the dimension $D$ is even. 
To show this, let $M_1$ be the largest absolute value of $\lambda_i$ ($M_1 := \mbox{max} _{i \in (1,\dots,D)} |\lambda _i|$), 
which is greater or equal to $|x|$ from Eq.~\eqref{eq:lambda>x}. 
Let $n_1$ and $m_1$ be the numbers of $\lambda_i$ satisfying $\lambda_i = M_1$ and $\lambda_i = - M_1$, respectively. 
If $M_1 > |x|$, then it can be shown that $n_1 = m_1$. 
Indeed we have from Eq.~(\ref{eq6})
\begin{equation}
 (n_1-m_1) + \sum_{i \in C_1} \bigg( \frac{\lambda_i}{M_1} \bigg)^{2k-1} = \bigg( \frac{x}{M_1}\bigg)^{2k-1}, 
\end{equation}
where $C_1 =$  \{\ $ i \in \{1,\ldots,D\} \ | \ |\lambda_i| < M_1$ \}\ .
By taking the limit $k \rightarrow \infty$, we obtain $n_1= m_1$, and Eq.~(\ref{eq6}) becomes 
\begin{equation}
 \sum_{i \in C_1} \lambda _{i}^{2k-1} = x^{2k-1}.
\end{equation}
Applying the same procedure for the second (the third, $\ldots$) largest absolute value $M_2 (M_3, \ldots )$ of $\lambda_i$, with 
$n_2 (n_3, \ldots )$ and $m_2(m_3, \ldots)$ being the numbers of $\lambda_i$ satisfying $\lambda_i = M_2 (M_3, \ldots)$ and $\lambda_i = - M_2 (-M_3, \ldots )$, we have $n_j = m_j$ as far as $M_j > |x| \ (j=2,3, \ldots)$ 
and Eq.~(\ref{eq6}) finally becomes 
\begin{equation}
 \sum_{i \in C} \lambda _{i}^{2k-1} = x^{2k-1}, 
\end{equation}
where $C = \{i \in \{1,\ldots,D\} \ | \ |\lambda_i| = |x| \}$. 
This requires that the numbers of $n_C$ and $m_C$ such that $\lambda_i = |x|$ and $\lambda_i = - |x|$ satisfy $n_C - m_C = 1$. 
To summarize, we have shown that $\vec \lambda$ has the following structure: 
\begin{eqnarray}
 \vec \lambda = (\underbrace{M_1,\dots}_{n_1}, \underbrace{-M_1,\dots}_{n_1}, \dots, \underbrace{M_j,\dots}_{n_j}, \underbrace{-M_j,\dots}_{n_j}, \nonumber \\ 
 \underbrace{|x|,\dots}_{n_C}, \underbrace{-|x|,\dots}_{n_C-1} ).
\end{eqnarray}
From this, we have 
\begin{equation}
 2 (\sum_{j} n_j + n_C) -1 = D, \label{eq9}
\end{equation}
which implies that the dimension $D$ is odd. 
By contradiction, we conclude that it is impossible to measure the purity of a quantum state (and thus the overlap between two quantum states) within the minimal model in the case of even dimensional quantum systems (the latter parts of Theorem \ref{thm:main} and Proposition \ref{prop:Overlap}).
\hfill $\blacksquare$

\section{Construction of a minimal model}\label{sec:ex}
In this section, for the reader's convenience, we give examples of a measurement operator and  a unitary operator in the minimal model of the purity measurement for odd dimensional cases (Let $D = 2N+1$). 
From the proof of Theorem \ref{thm:main}, the purity measurement in the minimal model is possible if a local PVM element of rank $N+1 (= (D+1)/2)$ is chosen as 
\begin{equation}
 P = \sum_{i=0}^{N} \ket{i}\bra{i}, 
\end{equation}
where \{\ $\ket{k}$ \}\ $_{k=0}^{2N}$ is an ONB of $\HA$. 

An example of a unitary operator $U$ necessary for the minimal model is given by 
\begin{equation}\label{eq:VU}
U = {\cal V}{\cal U},
\end{equation}
with ${\cal U} = \Pi_{i=0}^{2N-1}\Pi_{j=i+1}^{2N} U_{i,j} $ and ${\cal V} = \Pi_{i=0}^{N-1}\Pi_{j=i+1}^{N} V_{i,j} $ where 
\begin{equation}
U_{i,j}  =\exp \bigg[ \frac{\pi}{4}(\ket{i,j}\bra{j,i} - \ket{j,i}\bra{i,j})\bigg],
\end{equation}
\begin{equation}
\begin{split}
V_{i,j} = \exp \bigg[ &\frac{\pi}{2}(\ket{2N+1-j,2N+1-j+i}\bra{j,i} \\
            &- \ket{j,i}\bra{2N+1-j,2N+1-j+i}) \bigg].
\end{split} 
\end{equation}
Indeed, unitary gates $U_{i,j}(0\leq i<j\leq 2N)$ transform bases $\ket{i,j},\ket{j,i}, \ket{\mbox{other bases}}$ into
\begin{eqnarray}
 &&U_{i,j} \ket{i,j} = \frac{1}{\sqrt{2}}(\ket{i,j} -\ket{j,i}),  \\
 &&U_{i,j} \ket{j,i} = \frac{1}{\sqrt{2}}(\ket{i,j} +\ket{j,i}),  \\
 &&U_{i,j} \ket{\mbox{other bases}} = \ket{\mbox{other bases}},
\end{eqnarray}
and unitary gates $V_{i,j}(0\leq i<j\leq N)$ transform bases $\ket{j,i},\ket{2N+1-j,2N+1-j+i}, \ket{\mbox{other bases}}$ into
\begin{eqnarray}
 &&V_{i,j} \ket{j,i} = \ket{2N+1-j,2N+1-j+i}, \\
 &&V_{i,j} \ket{2N+1-j,2N+1-j+i} = -\ket{j,i}, \\
 &&V_{i,j} \ket{\mbox{other bases}} = \ket{\mbox{other bases}}.
\end{eqnarray}
Hence, the relation between the swap operator and the PVM element is given by
\begin{equation}
USU^{\dagger} = - \I \otimes \I + 2 P \otimes \I,
\end{equation}
and thus one can estimate the purity within the minimal model as $P(\rho) = 2 \mbox{Pr}(\rho) -1$.

In the simplest case, i.e., in the case of qutrit ($D=3$), the final projection operator $P$ is described by
\begin{equation}
P = \begin{pmatrix} 1 && \\ &1& \\ && 0 \end{pmatrix},
\end{equation}
in a basis \{\ $\ket{0} , \ket 1 , \ket 2$ \}\ and a unitary operator $U$ necessary for the minimal model is given by
\begin{equation}
U = e^{-i \frac{\pi}{2}H_4} e^{-i \frac{\pi}{4}H_3} e^{-i \frac{\pi}{4}H_2} e^{-i \frac{\pi}{4}H_1},
\end{equation}
where
\begin{equation}
H_1 = i \begin{pmatrix} 0 &1& 0 \\ 0 &0& 0 \\ 0 &0& 0 \end{pmatrix} \otimes \begin{pmatrix} 0 &0& 0 \\ 1 &0& 0 \\ 0 &0& 0 \end{pmatrix} - i \begin{pmatrix} 0 &0& 0 \\ 1 &0& 0 \\ 0 &0& 0 \end{pmatrix} \otimes \begin{pmatrix} 0 &1& 0 \\ 0 &0& 0 \\ 0 &0& 0 \end{pmatrix},
\end{equation}
\begin{equation}
H_2 = i\begin{pmatrix} 0 &0& 1 \\ 0 &0& 0 \\ 0 &0& 0 \end{pmatrix} \otimes \begin{pmatrix} 0 &0& 0 \\ 0 &0& 0 \\ 1 &0& 0 \end{pmatrix} - i\begin{pmatrix} 0 &0& 0 \\ 0 &0& 0 \\ 1 &0& 0 \end{pmatrix} \otimes \begin{pmatrix} 0 &0& 1 \\ 0 &0& 0 \\ 0 &0& 0 \end{pmatrix},
\end{equation}
\begin{equation}
H_3 = i\begin{pmatrix} 0 &0& 0 \\ 0 &0& 1 \\ 0 &0& 0 \end{pmatrix} \otimes \begin{pmatrix} 0 &0& 0 \\ 0 &0& 0 \\ 0 &1& 0 \end{pmatrix} - i\begin{pmatrix} 0 &0& 0 \\ 0 &0& 0 \\ 0 &1& 0 \end{pmatrix} \otimes \begin{pmatrix} 0 &0& 0 \\ 0 &0& 1 \\ 0 &0& 0 \end{pmatrix},
\end{equation}
\begin{equation}
H_4 = i\begin{pmatrix} 0 &0& 0 \\ 0 &0& 0 \\ 0 &1& 0 \end{pmatrix} \otimes \begin{pmatrix} 0 &0& 0 \\ 0 &0& 0 \\ 1 &0& 0 \end{pmatrix} - i\begin{pmatrix} 0 &0& 0 \\ 0 &0& 1 \\ 0 &0& 0 \end{pmatrix} \otimes \begin{pmatrix} 0 &0& 1 \\ 0 &0& 0 \\ 0 &0& 0 \end{pmatrix}.
\end{equation}

In Appendix \ref{sec:even_case}, we give examples of a measurement operator and a unitary operator for purity measurement using 
a local two-body measurement (See Remark \ref{rem:LocalTwoBody}).  

\section{Summary and discussion}\label{s_and_d}
In this paper, we have investigated the possibility of measuring the purity of any unknown quantum state
(and the overlap between two quantum states) within a minimal measurement scheme such that 
all elements consisting of the device are a unitary interaction and a local yes-no measurement on a single system without using any ancilla systems or random measurements.
We have shown that the measurability of the purity and the overlap within the minimal model critically depends on
the parity of dimension of quantum systems: for odd (and infinite) dimensional cases, it is possible 
while it is impossible for even dimensional cases. 

This difference results from two restrictions on the minimal purity-measurement scheme: the transformation of a bipartite system is a unitary transformation and the final measurement is a local measurement on a single system. Roughly speaking, since the bipartite system is unitarily transformed, the projection operator of the final measurement must be chosen as the projector onto the symmetric or antisymmetric subspace of the swap operator. It is, however, possible only if the dimension of Hilbert space associated to a quantum system is odd as we have proved. For example, in the case of qubit ($D=2$), since the dimensions of symmetric and antisymmetric subspaces are $3$ and $1$, respectively, the rank of the final projector must be $3$ or $1$. However, the rank of any local projector $P \otimes \I$ cannot be chosen as this. On the other hand, in the case of qutrit ($D=3$), the dimensions of symmetric and antisymmetric subspaces are $6$ and $3$, respectively, and there exists a local projector $P \otimes \I$ of the rank $3$ or $6$.


The dimension of Hilbert space associated to a quantum system is of both fundamental and applicational interest in quantum (information) theory (See e.g., Refs.~\cite{ref:Fuchs,determine_dim} and references therein). 
The quantity can have a physical/operational meaning such as the maximum number of distinguishable states. 
The intriguing point is that there appears individual character for each dimension. 
For instance, Kochen-Specker theorem \cite{hidden} tells us that there exists a non-contextual hidden variable model iff the dimension is two; 
The geometry of state space is affinely isomorphic to a ball iff the dimension is two \cite{kimura}; The existence of the complete mutually unbiased bases (MUBs) is known only for the power prime dimensional cases \cite{ref:WF} (while it is believed that there are dimensions without complete MUBs). 
From the applicational point of view, the dimension plays a crucial role in security proofs of standard quantum key distribution schemes \cite{qkb}. 
We think that our result would also be regarded as an example of physical difference between odd and even dimensional quantum systems.

\acknowledgments

We would like to thank Professor S. Pascazio, Professor K. Yuasa, Professor M. Hayashi, Dr. M. Unoki, and Mr. M. Kanazawa for useful comments and advice.
This work was partly supported by a Grant-in-Aid for Young Scientists (B) (No. 22740079) and a Grant-in-Aid for Scientific Research (C) (No. 22540292) from JSPS. 
\appendix

\section{Linearity of function}\label{sec:linearity}
We give the proof of Proposition \ref{prop:linearity} focusing on the degree of parameters	 which characterize a density operator $\rho$.
In terms of the generators $\{T^i\}_{i=1}^{D^2-1}$ of $SU(D)$, let us first expand the density operator $\rho$ as
\begin{equation}
 \rho = \frac{1}{D}\I + \sqrt{\frac{D-1}{2D}} \sum_{i=1}^{D^2-1}a_i T^i,
\end{equation}
where $a_i$ is a component of a generalized Bloch vector (See e.g. \cite{kimura}).
Differentiating both sides of Eq.~(\ref{eq:cond}) with respect to $a_k$, we have
\begin{equation}
\frac{\partial {P}(\rho)}{\partial a_k} = \frac{\partial f(\mbox{Pr}(\rho))}{\partial (\mbox{Pr}(\rho))}\frac{\partial (\mbox{Pr}(\rho))}{\partial a_k}. \label{p1}
\end{equation}
Since
\begin{equation}
P(\rho) = \frac{1}{D} + \frac{D-1}{D}\sum_{i=1}^{D^2-1}a_i^2 ,
\end{equation}
we have
\begin{equation}
\frac{\partial P(\rho)}{\partial a_k} = \frac{2(D-1)}{D}a_k. \label{p2}
\end{equation}
From Eq.~(\ref{p2}),  the degree of the parameter $a_k$ on both sides of Eq.~(\ref{p1}) is consistent when either of the following conditions holds for all $k \in$ \{\ $1,\dots,N^2 -1$ \}\ :
\begin{equation}
a_k \in \frac{\partial (\mbox{Pr}(\rho))}{\partial a_k}, \ a_k \not \in \frac{\partial f(\mbox{Pr}(\rho))}{\partial (\mbox{Pr}(\rho))},\label{p4}
\end{equation}
\begin{equation}
a_k \not \in \frac{\partial (\mbox{Pr}(\rho))}{\partial a_k}, \ a_k \in \frac{\partial f(\mbox{Pr}(\rho))}{\partial (\mbox{Pr}(\rho))}.
\end{equation}
In the latter case, however, the function $f$ explicitly has the dependence on the parameter $a_k$, which means besides the probability $\rm{Pr}(\rho)$
additional information about the parameter $a_k$ is neccesary to measure the purity.  
Therefore, Eq.~(\ref{p4}) is the only consistent condition and we finally have
\begin{equation}
^{\exists }a\not =0 \in \R \ {\rm s.t.} \frac{\partial f(\mbox{Pr}(\rho))}{\partial (\mbox{Pr}(\rho))} = a, \ \forall \rho \in \SA(\HA),
\end{equation}
and thus
\begin{equation}
\exists a \not= 0,b \in \R \ {\rm s.t.}  f(\mbox{Pr}(\rho)) =  a \mbox{Pr}(\rho) + b, \ \forall \rho \in \SA(\HA).
\end{equation}\hfill $\blacksquare$

\section{Local yes-no measurement for even dimensional case}\label{sec:even_case}

We give examples of a measurement operator and a unitary operator for measuring the purity with a local two-body measurement $P \otimes Q$ for even dimensional cases (Let $D = 2N$). 
From the proof of Theorem \ref{thm:main}, one can choose the measurement operator $P \otimes Q$ as
\begin{equation}
P \otimes Q = \bigg( \sum_{i=0}^{N-1} \ket{i} \bra{i} \bigg) \otimes \bigg( \sum_{i=0}^{2N-2} \ket{i} \bra{i} \bigg),
\end{equation}
where \{\ $\ket{k}$ \}\ $_{k=0}^{2N-1}$ is an ONB of $\HA$.

Compared with Eq.~\eqref{eq:VU}, the construction of a unitary operator is slightly intricate: 
\begin{equation}
U = {\cal T} {\cal W} {\cal R},
\end{equation}
with ${\cal R} = \Pi_{i=0}^{2N-1} \Pi_{j=i+1}^{2N} R_{i,j}$, ${\cal T} = \Pi_{i=0}^{N-2}\Pi_{j=i+1}^{N-1} T_{i,j}$ and ${\cal W} = \Pi_{i=0}^{N-1} W_{i}$ where
\begin{equation}
R_{i,j} = U_{i,j}^{\dagger} = \exp \bigg[ \frac{\pi}{4} (\ket{j,i} \bra{i,j} - \ket{i,j} \bra{j,i} ) \bigg],
\end{equation}
\begin{equation}
T_{i,j} = \exp \bigg[ \frac{\pi}{2} (\ket{j,i} \bra{N+i,N+j} - \ket{N+i,N+j} \bra{j,i} ) \bigg],
\end{equation}
\begin{equation}
W_{i} = \exp \bigg[ \frac{\pi}{2} (\ket{i,i} \bra{i,2N-1} - \ket{i,2N-1} \bra{i,i} ) \bigg] .
\end{equation}
Indeed, the gates $R_{i,j} (0 \leq i < j \leq 2N-1)$ transform bases $\ket{i,j}$, $\ket{j,i}$ and $\ket{\mbox{other bases}}$ into
\begin{eqnarray}
 &&R_{i,j} \ket{i,j} = \frac{1}{\sqrt{2}}(\ket{i,j} +\ket{j,i}), \\
 &&R_{i,j} \ket{j,i} = \frac{1}{\sqrt{2}}(-\ket{i,j} +\ket{j,i}), \\
 &&R_{i,j} \ket{\mbox{other bases}} = \ket{\mbox{other bases}},
\end{eqnarray}
the gates $T_{i,j} (0 \leq i < j \leq N-1)$ transform bases $\ket{j,i}$, $\ket{N + i,N+j}$ and $\ket{\mbox{other bases}}$ into
\begin{eqnarray}
 &&T_{i,j} \ket{j,i} = -\ket{N+i,N+j}, \\
 &&T_{i,j} \ket{N+i,N+j} = \ket{j,i}, \\
 &&T_{i,j} \ket{\mbox{other bases}} = \ket{\mbox{other bases}},
\end{eqnarray}
and the gates $W_{i} (0 \leq i \leq N-1)$ transform bases $\ket{i,i}$, $\ket{i,2N-1}$ and $\ket{\mbox{other bases}}$ into
\begin{eqnarray}
 &&W_{i} \ket{i,i} = -\ket{i,2N-1}, \\
 &&W_{i} \ket{i,2N-1} = \ket{i,i}, \\
 &&W_{i} \ket{\mbox{other bases}} = \ket{\mbox{other bases}}.
\end{eqnarray}
Hence, the relation between the swap operator and the PVM element is given by
\begin{equation}
USU^{\dagger} =  \I \otimes \I - 2 P \otimes Q,
\end{equation}
and thus one can estimate the purity with the local two-body measurement as $P(\rho) = 1- 2\mbox{Pr}(\rho)$.

In the simplest case, i.e., in the case of qubit ($D=2$), the final projection operator $P \otimes Q$ is described by
\begin{equation}
P \otimes Q = \begin{pmatrix} 1& \\ &0 \end{pmatrix} \otimes \begin{pmatrix} 1& \\ &0 \end{pmatrix},
\end{equation}
in a basis \{\ $\ket{0} , \ket 1$ \}\ and a unitary operator $U$ necessary for the local yes-no measurement is given by
\begin{equation}
U = e^{-i \frac{\pi}{2}H_2} e^{-i \frac{\pi}{4}H_1},
\end{equation}
where
\begin{equation}
H_1 = i\begin{pmatrix} 0 & 0 \\ 1 & 0 \end{pmatrix} \otimes \begin{pmatrix} 0 & 1 \\ 0 & 0 \end{pmatrix} - i\begin{pmatrix} 0 & 1 \\ 0 & 0 \end{pmatrix} \otimes \begin{pmatrix} 0 & 0 \\ 1 & 0 \end{pmatrix},
\end{equation}
\begin{equation}
H_2 = i\begin{pmatrix} 1 & 0 \\ 0 & 0 \end{pmatrix} \otimes \begin{pmatrix} 0 & 1 \\ 0 & 0 \end{pmatrix} - i\begin{pmatrix} 1 & 0 \\ 0 & 0 \end{pmatrix} \otimes \begin{pmatrix} 0 & 0 \\ 1 & 0 \end{pmatrix}.
\end{equation}


\begin{thebibliography}{99}

\bibitem{ref:vN}
J. von Neumann, {\em Mathematische Grundlagen der Quantenmechanik} (Springer, Berlin, 1932).

\bibitem{ref:OptimalRate}
B. Schumacher, Phys. Rev. A {\bf 51} 2738 (1995). 
\bibitem{ref:EntDis}
C. H.~Bennett, H. J.~Bernstein, S. Popescu, B. Schumacher, Phys. Rev. A {\bf 53} 2046 (1996).
\bibitem{NC}
M. A. Nielsen and I. L. Chuang, {\em Quantum Computation and Quantum Information\/} (Cambridge University Press, Cambridge, 2000).

\bibitem{ref1}
{\it The Physics of Quantum Information\/}, edited by D. Bouwmeester, A. Ekert, and A. Zeilinger (Springer-Verlag, Berlin, 2000).

\bibitem{ref2}
M. Namiki, S. Pascazio, and H. Nakazato, {\it Decoherence and Quantum Measurements\/} (World Scientific, Singapore, 1997).

\bibitem{ref:WoottersConc}
S. Hill and W. K.~Wootters, Phys. Rev. Lett. {\bf 78} 5022 (1997); 
W. K.~Wootters, Quantum Inf.\ Comp.\ \textbf{1} , 27 (2001). 
\bibitem{ref:Conc}
V. Coffman, J. Kundu, and W. K. Wootters, Phys. Rev. A {\bf 61}, 052306 (2000).
\bibitem{tomography}
M. A. Nielsen and I. L. Chuang, {\it Quantum Computation and Quantum Information\/} (Cambridge University Press, Cambridge, 2000).

\bibitem{ref:flip} 
R. Filip, Phys.\ Rev.\ A {\bf65}, 062320 (2002).

\bibitem{ref:ekert} 
A. K. Ekert, C. M. Alves, D. K. L. Oi, M. Horodecki, P. Horodecki, and L. C. Kwek, Phys.\ Rev.\ Lett.\ {\bf 88}, 217901 (2002).

\bibitem{ref:NTYFP}
H. Nakazato, T. Tanaka, K. Yuasa, G. Florio, and S. Pascazio, Phys. Rev. A {\bf 85}, 042316 (2012).


\bibitem{random}
S. J. van Enk and C. W. J. Beenakker, Phys.\ Rev.\ Lett.\ {\bf 108}, 110503 (2012).

\bibitem{ref:Brun}
T. A. Brun, 
Quantum Inf.\ Comp.\ \textbf{4}, 401 (2004).


\bibitem{ref:Dirac}
P. A.~M. Dirac, {\em The Principles of Quantum Mechanics} (Oxford, 1958).

\bibitem{ref:Sch}
R. Schatten, {\em Norm Ideals of Completely Continuous Operators} (Springer-Verlag, Berlin, 1960). 


\bibitem{ref:Fuchs}
C.~A. Fuchs,
Quantum Inf.\ Comp.\ \textbf{4}, 467 (2004).

\bibitem{determine_dim}
M. Hendrych, R. Gallego, M. Mi${\rm \check c}$uda, N. Brunner, A. Ac\'in, J. P. Torres, Nature Phys.\ {\bf 8}, 588 (2012). 


\bibitem{hidden}
S. Kochen, E. P. Specker, J.\ Math.\ Mech.\ {\bf 17}, 59 (1967);
A. Peres, J.\ Phys.\ A: Math.\ Gen.\ {\bf 24}, L175 (1991).

\bibitem{kimura}
G. Kimura, Phys.\ Lett.\ A {\bf314}, 339 (2003); 
M. S. Byrd, N. Khaneja, Phys. Rev. A {\bf 68} 062322 (2003). 


\bibitem{ref:WF}
W. K. Wootters and B. C. Fields, Ann. Phys. {\bf 191}, 363 (1989). 

\bibitem{qkb}
A. Ac\'in, N. Gisin, L. Masanes, Phys.\ Rev.\ Lett.\ {\bf 97}, 120405 (2006).
\end{thebibliography}
\end{document}